\documentclass{article} % For LaTeX2e
\usepackage{iclr2026_re-align_workshop,times}

\usepackage{amsmath,amsfonts,bm}

\def\eqref#1{equation~\ref{#1}}
\def\1{\bm{1}}

\DeclareMathAlphabet{\mathsfit}{\encodingdefault}{\sfdefault}{m}{sl}
\SetMathAlphabet{\mathsfit}{bold}{\encodingdefault}{\sfdefault}{bx}{n}

\usepackage[hidelinks]{hyperref}
\usepackage{url}
\usepackage[pdftex]{graphicx} %%%% added
\usepackage{seqsplit}       % Useful for splitting very long strings of code
\usepackage{natbib}         % Required for your \citep commands
\usepackage[T1]{fontenc}
\usepackage[utf8]{inputenc}

\iclrfinalcopy
\title{Modulating Cross-Modal Convergence with Single-Stimulus, Intra-Modal Dispersion}

\author{Eghbal A. Hosseini\textsuperscript{1}\thanks{Now at Google DeepMind} , Brian Cheung\textsuperscript{2}, Evelina Fedorenko\textsuperscript{1} \& Alex H. Williams\textsuperscript{3,4} \\
\textsuperscript{1}Department of Brain and Cognitive Sciences, MIT \\
\textsuperscript{2}CSAIL, MIT \\
\textsuperscript{3}Center for Neural Science, NYU \\
\textsuperscript{4}Center for Computational Neuroscience, Flatiron Institute \\
\texttt{\{ehoseini, evelina9\}@mit.edu, cheungb@mit.edu, aw4614@nyu.edu} \\
}

\begin{document}

\maketitle

\begin{abstract}
Neural networks exhibit a remarkable degree of representational convergence across diverse architectures, training objectives, and even data modalities. This convergence is predictive of alignment with brain representation. A recent hypothesis suggests this arises from learning the underlying structure in the environment in similar ways. However, it is unclear how individual stimuli elicit convergent representations across networks. An image can be perceived in multiple ways and expressed differently using words. Here, we introduce a methodology based on the Generalized Procrustes Algorithm to measure intra-modal representational convergence at the single-stimulus level. We applied this to vision models with distinct training objectives, selecting stimuli based on their degree of alignment (intra-modal dispersion). Crucially, we found that this intra-modal dispersion strongly modulates alignment between vision and language models (cross-modal convergence). Specifically, stimuli with low intra-modal dispersion (high agreement among vision models) elicited significantly higher cross-modal alignment than those with high dispersion, by up to a factor of two (e.g., in pairings of DINOv2 with language models). This effect was robust to stimulus selection criteria and generalized across different pairings of vision and language models. Measuring convergence at the single-stimulus level provides a path toward understanding the sources of convergence and divergence across modalities, and between neural networks and human neural representations.
\end{abstract}

\section{Introduction}

Artificial neural networks show a remarkable ability to learn representations that generalize across tasks, and predict neural representations in humans and other animals. Even though details of implementation — including training, architecture, and input modality — vary between networks, they appear to have minimal influence on the final representation, and models converge onto similar representations. The Platonic Representation Hypothesis \citep{huh2024platonicrepresentationhypothesis} suggests this convergence arises from learning shared environmental priors. \cite{Hosseini2024} found similar evidence in convergence across neural networks and their predictivity of brain representation. Both lines of work however established convergences across groups of stimuli, and thus globally.

Individual observations are critical for probing and understanding representations and can potentially drive models, and humans, toward either convergence or divergence. Humans often find themselves interpreting differently the same work of art, a painting for example.
Existing methods are often ill-suited for probing convergence at single stimulus level, as they either measure local effects \citep{Feather2023-dj}, average similarity over a large set of stimuli datasets \citep{Hosseini2024}, or lack a proper metric space for rigorous comparison \citep{sucholutsky2024gettingalignedrepresentationalalignment, harvey2023dualityburesshapedistances}. This highlights the need for a robust, stimulus-specific measure of alignment. 

Building on Generalized Procrustes Analysis \citep{gower1975generalized}, we introduce a stimulus-specific measure of representational convergence across vision models \citep{williams2021generalized,10.7554/eLife.56601}. We then demonstrate that our measure can be used to select stimuli that effectively modulate representational alignment across modalities, from vision to language.

\section{Methods}
\subsection{Generalized Procrustes method}

Generalized Procrustes Analysis (GPA) is a method for aligning multiple datasets, or embedding spaces, into a common reference frame. The core idea is to find a set of optimal transformations, restricted to orthogonal transformations, that minimize the discrepancy between a shared consensus configuration and each individually transformed dataset \citep{gower1975generalized}. Stimulus-level residuals are a classical by-product of GPA; our contribution is their application to comparing neural network representations and to modulating cross-modal alignment.

Formally, consider a set of $M$ representation matrices $\{N_1, N_2, \dots, N_M\}$, where each representation $N_i \in \mathbb{R}^{m \times n_i}$, with $i \in \{1, \dots, M\}$ indexing models, represents $m$ corresponding samples (e.g., stimuli) in an $n_i$-dimensional space. The dimensionality $n_i$ can differ across the representations.
GPA seeks to find an optimal consensus representation, $N_{\text{joint}} \in \mathbb{R}^{m \times k}$, and a set of corresponding orthogonal transformation matrices, $\{T_1, T_2, \dots, T_M\}$, where each $T_i \in \mathbb{R}^{n_i \times k}$, by minimizing the sum of squared Frobenius distances. The optimization problem is defined as:
\begin{equation}
\min_{N_{\text{joint}}, \{T_i\}_{i=1}^M} \sum_{i=1}^{M} \| N_i T_i - N_{\text{joint}} \|_F^2
\end{equation}
subject to each transformation matrix being orthogonal, $
T_i^T T_i = I_k$ for all $i \in \{1, \dots, M\}$.
Here, $\| \cdot \|_F$ denotes the Frobenius norm, and the constraint $T_i^T T_i = I_k$ ensures that the transformations do not distort the internal geometry of each representation $N_i$.
This process effectively rotates each embedding space to achieve maximal alignment with the emergent consensus space $N_{\text{joint}}$.
Prior to performing GPA, we first center each embedding along the second dimension $n_i$, zero-pad dimensions across all models to a common dimensionality, and normalize each representation $\|N_i\|_F=1$. 

Generalized Procrustes Analysis can also be interpreted as computing a \textbf{barycenter} of neural representations in Procrustes shape space. In classical shape analysis, each centered, scale-normalized configuration $N_i$ corresponds to a point on a quotient manifold obtained by modding out rotations,\footnote{Here we include reflections as well as rotations in the equivalence class. This is common in the analysis of neural representations since even a permutation of the neural unit labels can result in a reflection.} and Generalized Procrustes iteratively aligns all configurations to the barycenter (also called the Fr\'echet mean or Karcher mean), represented by $N_{\text{joint}}$.
This perspective clarifies that GPA is not merely an arbitrary linear alignment, but a principled procedure for finding the central tendency of a population of neural representations with respect to a scale and rotation-invariant metric space \citep{williams2021generalized}.

We restrict each $T_i$ to orthogonal transformations, rather than a general linear map, because they preserve the internal geometry of each representation: pairwise distances and angles between stimuli are unchanged under $T_i$. A general linear map can absorb scale and reshape covariance structure, which are not desirable for defining metrics on representation \citep{williams2021generalized}. Scale is instead handled explicitly by the Frobenius normalization $\|N_i\|_F = 1$.

\subsection{Quantifying Single-Stimulus Dispersion}
\begin{figure}
  \centering
  \includegraphics[width=5.25in]{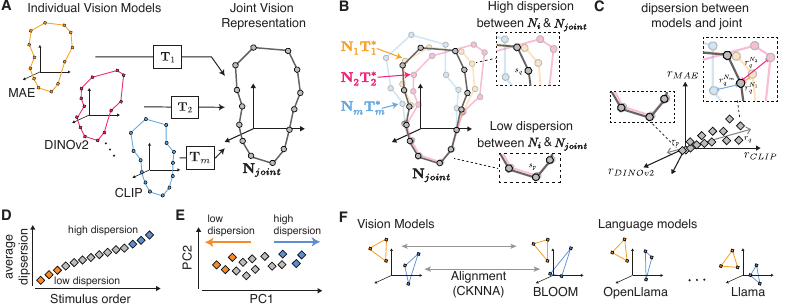}
  \caption{\textbf{(A)} Overview of the Generalized Procrustes Analysis (GPA) problem. Given representations from diverse vision models, our goal is to learn a set of model-specific transformations to construct a single joint representation. \textbf{(B)} When individual model representations are projected onto this joint space, stimuli can exhibit either low dispersion between joint and individual models (bottom) or instead high dispersion (top). \textbf{(C)} This dispersion is quantified in a residual space, where each stimulus is represented as a point whose coordinates reflect its distance between each model's representation and the joint space. \textbf{(D, E)} We use two approaches to identify stimuli with varying degrees of dispersion: (D) ranking stimuli based on their average dispersion across all models and (E) using the score along the first principal component (PC1) of the residual space as a more robust measure. \textbf{(F)} After identifying stimuli with high and low dispersion, we measure the alignment between the vision models and language models using a local similarity metric (CKNNA) to test how stimulus selection modulates cross-modal convergence.}
  \label{fig:methods}
\end{figure}

To quantify representational convergence at the single-stimulus level, we measure the dispersion between each model's transformed representation and the shared joint representation, $N_{\text{joint}}$. Specifically, after computing the joint space via GPA, we first project each model's representation into this common space using its corresponding transformation matrix, $T_i$. For each stimulus $j \in \{1, \dots, m\}$, we then compute the Euclidean distance between its projected representation from model $i$ and its joint representation. This yields the \textbf{Procrustes residual}, $r_{ij}$, defined as:
\begin{equation}
r_{ij} = \| (N_i T_i)_j - (N_{\text{joint}})_j \|_2
\label{eq:residual}
\end{equation}
This process results in a residual matrix $R \in \mathbb{R}^{m \times M}$, where each entry $r_{ij}$ quantifies the dispersion for a given stimulus $j$ and model $i$ (see Fig~\ref{fig:methods}C). A low residual value signifies high alignment (low dispersion), while a high value indicates low alignment (high dispersion).

We employ two strategies to summarize the overall dispersion for each stimulus across all $M$ models:

\begin{enumerate}
    \item \textbf{Mean Dispersion:} Our first approach is to average the residuals across all models for each stimulus. We rank the stimuli from least to most dispersed. While straightforward, this metric can be sensitive to outlier models (Fig~\ref{fig:methods}D)

    \item \textbf{Principal Component of Dispersion:} To capture the primary axis of disagreement in a more robust manner, our second approach utilizes Principal Component Analysis (PCA) on the residual matrix $R$. We use the score along the first principal component (PC1) as a more nuanced measure of dispersion for each stimulus, as it reflects the most significant shared variance in model disagreement (Fig~\ref{fig:methods}E).
\end{enumerate}

\subsection{Vision Models}

% We inspected representations across ViT models trained on diverse objectives \citep{huh2024platonicrepresentationhypothesis}. These objectives include Masked Autoencoders (MAE) \citep{He_2022_CVPR}, DINO \citep{Caron_2021_ICCV}, CLIP \citep{pmlr-v139-radford21a}, and CLIP with additional finetuning on ImageNet-12K. To compute the joint Procrustes representation, we selected a large architecture in each class: MAE: \texttt{vit\_huge\_patch14\_224.mae}; DINO: \texttt{vit\_giant\_patch14\_dinov2.lvd142m}; CLIP: \texttt{vit\_huge\_patch14\_clip\_224.laion2b}; CLIP + finetuning on ImageNet-12K: \texttt{vit\_huge\_patch14\_clip\_224.laion2b\_ft\_in12k}. All ViTs have a dimension of 1280, except for the DINO model, which had a dimension of 1536. We thus zero-padded all other model representations to match the DINO dimensions. We extracted the CLS-token representation from the penultimate transformer block of each ViT and used it for Procrustes alignment, and identified stimuli with varying degrees of dispersion.

We inspected representations across ViT models trained on diverse objectives \citep{huh2024platonicrepresentationhypothesis}. These objectives include Masked Autoencoders (MAE) \citep{He_2022_CVPR}, DINO \citep{Caron_2021_ICCV}, CLIP \citep{pmlr-v139-radford21a}, and CLIP with additional finetuning on ImageNet-12K. To compute the joint Procrustes representation, we selected a large architecture in each class: 
MAE: \texttt{vit\_huge\_patch14\_224.mae}; 
DINO: \texttt{vit\_giant\_patch14\_dinov2.lvd142m}; 
CLIP: \texttt{vit\_huge\_patch14\_clip\_224.laion2b}; 
CLIP + finetuning on ImageNet-12K: \texttt{vit\_huge\_patch14\_clip\_224.laion2b\_ft\_in12k}. 

All ViTs have a dimension of 1280, except for the DINO model, which had a dimension of 1536. We thus zero-padded all other model representations to match the DINO dimensions. We extracted the CLS-token representation from the penultimate transformer block of each ViT and used it for Procrustes alignment, and identified stimuli with varying degrees of dispersion.

\subsection{Language Models}

Similar to \cite{huh2024platonicrepresentationhypothesis} we compared representations of ViT vision model to LLMs across model families. We focused on BLOOM \citep{workshop2023bloom176bparameteropenaccessmultilingual} , OpenLLaMA \citep{openlm2023openllama}, and LLaMA model class \citep{touvron2023llama2openfoundation}. For each LLM, we extracted token embeddings from each transformer block and mean-pooled across tokens to obtain a single representation per caption, we then used the block that showed best alignment with the vision models.

\subsection{Dataset}
Following \cite{huh2024platonicrepresentationhypothesis}, we used Wikipedia caption dataset \citep{10.1145/3404835.3463257} to measure convergence between modalities. This dataset included a set of 1024 image/caption pairs from Wikipedia articles. Images and captions are used as released in the Wikipedia caption dataset (WIT); no additional filtering was applied.

\subsection{Alignment measure}
To measure the local alignment between vision and language modalities, we employed Centered Kernel k-Nearest Neighbors Alignment (CKNNA) \citep{huh2024platonicrepresentationhypothesis, pmlr-v97-kornblith19a}. This method adapts the standard Centered Kernel Alignment (CKA) to focus on local representational structure. Intuitively, CKNNA computes the cross-covariance only between a sample and its nearest neighbors, thereby assessing the local, rather than global, similarity between two representation spaces. We considered 10 nearest neighbors for each sample when reporting alignment (Fig.~\ref{fig:dispersion_results}).

\section{Results}

We investigated the dependence of convergence between visual and linguistic representations on stimuli using the Procrustes-based methodology described above. As a stricter test of the Platonic Representation Hypothesis, we considered representations learned across ViT transformer architectures with different objectives, an effect not explored in the original work. After computing a joint representation across these vision models, we extracted stimuli with varying degrees of dispersion to test their impact on cross-modal alignment.

\begin{figure}
  \centering
  \includegraphics[width=5.25in]{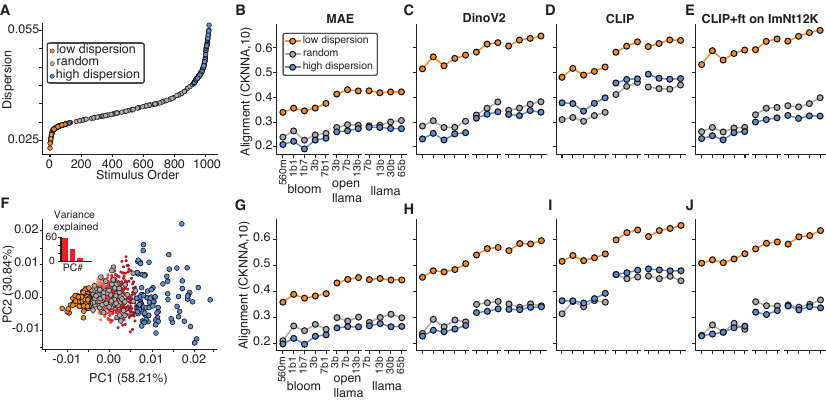}
   \caption{\textbf{Stimulus-specific dispersion modulates vision-language alignment.}
     \textbf{(A)} rank based stimulus selection:stimuli are sorted by mean dispersion to create low-, high-, and random-dispersion sets. \textbf{(B-E)} Vision-language alignment (CKNNA) is then measured for each set across four vision models (ViT-MAE, DINOv2, CLIP, and CLIP+FT).
    \textbf{(F)} PCA based stimulus selection: stimuli are selected based on their score along the first principal component (PC1) of the dispersion space. \textbf{(G-J)} The alignment experiment is repeated using these sets.
    For both selection methods, low-dispersion stimuli consistently yield the highest vision-language alignment across all tested models.}
    \label{fig:dispersion_results}
\end{figure}

Stimuli with low intra-modal dispersion yielded a substantially higher degree of cross-modal convergence between visual and linguistic representations. We first identified stimuli with the least and most dispersion using a rank-ordering of their mean residuals (Fig.~\ref{fig:dispersion_results}A). We selected three sets: (1) \textbf{low-dispersion}, (2) \textbf{high-dispersion}, and (3) \textbf{random baseline}. We then compared vision model alignment with three classes of LLMs. The low-dispersion set showed significantly higher convergence, up to a twofold increase in some cases (DinoV2, CLIP+ft on ImN12K), over the random and high-dispersion stimuli (Fig.~\ref{fig:dispersion_results}B-E).

The correspondence between intra-modal dispersion and cross-modal convergence generalizes to different sampling strategies. Instead of rank ordering, we performed PCA on the residual space and selected stimuli along the first principal component (PC1), as shown in (Fig.~\ref{fig:dispersion_results}F). When we repeated the alignment experiment, we found a consistent pattern where the degree of intra-modal dispersion strongly correlated with cross-modal convergence (Fig.~\ref{fig:dispersion_results}G-J). This confirms that the significant gap in alignment between low- and high-dispersion stimuli is a robust phenomenon across all tested models.

\section{Discussion}
We applied Generalized Procrustes Analysis to identify stimuli with varying representational dispersion across vision models at the single-stimulus level. We showed that selecting for low-dispersion stimuli can boost cross-modal alignment with language models close to twofold, and is robust across different selection criteria.

Our approach helps uncover what features drive representational convergence and could potentially offer a more stringent test for comparing artificial and biological neural networks. Future work could extend these findings to more datasets and, critically, identify the common features within low- and high-dispersion stimuli that are responsible for modulating alignment.

\bibliography{ICLR2026_realign_dispersion}
\bibliographystyle{iclr2026_conference}

% \newpage
% \appendix
% % Reset the figure counter to 0
% \setcounter{figure}{0}

% % Option A: Change the label to "Supplementary Figure 1"
% \renewcommand{\figurename}{Supplementary Figure}
% \section*{Supplementary Material}

\clearpage % Ensure the supplement starts on a new page
\section*{Supplementary Material}

% Reset figure counter and change prefix to 'S'
\setcounter{figure}{0} 
\renewcommand{\thefigure}{S\arabic{figure}}
\renewcommand{\figurename}{Supplementary Figure}

\begin{figure}[ht] 
  \centering
  \includegraphics[width=5.25in]{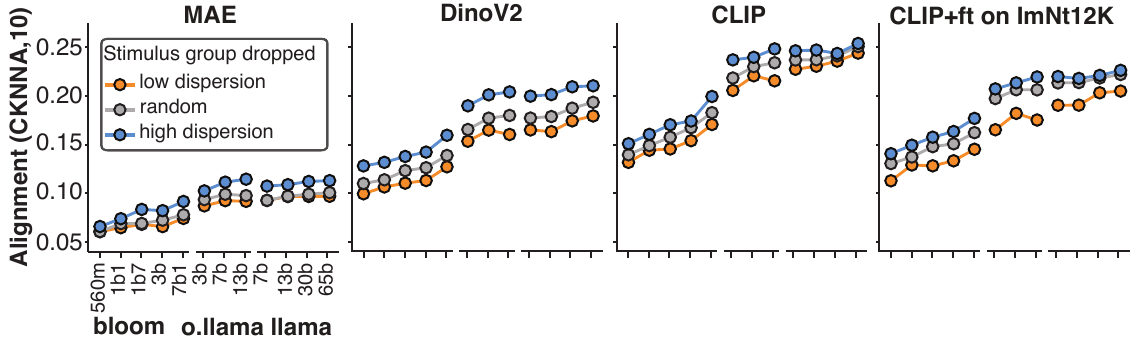}
  \caption[Effect of excluding dispersion on vision--language alignment]{\textbf{Excluding the highest-dispersion stimuli systematically increases vision-to-language alignment, whereas excluding the lowest-dispersion stimuli decreases it.} 
  We removed the 25\% of stimuli with the highest mean dispersion (\textit{drop high-dispersion}, blue) or the 25\% with the lowest (\textit{drop low-dispersion}, orange), and recomputed the vision--language local alignment (CKNNA, $k = 10$) on the remaining 768 stimuli from WIT. The full-set baseline (\textit{full}, gray) is shown for reference. Each of the four panels corresponds to one vision model (MAE, DINOv2, CLIP, CLIP+FT on ImageNet-12K), with alignment reported against 12 language models from three families: 
  BLOOM (\texttt{560m}, \texttt{1b1}, \texttt{1b7}, \texttt{3b}, \texttt{7b1}), OpenLLaMA (\texttt{3b}, \texttt{7b}, \texttt{13b}), and LLaMA (\texttt{7b}, \texttt{13b}, \texttt{30b}, \texttt{65b}). Across all vision-to-LLM pairings, excluding high-dispersion stimuli increases CKNNA above the full-set baseline, while excluding low-dispersion stimuli lowers it. \label{fig:S1_stimulus_dropping_fix}}
\end{figure}

\begin{figure}[h]

  \centering

  \includegraphics[width=5.25in]{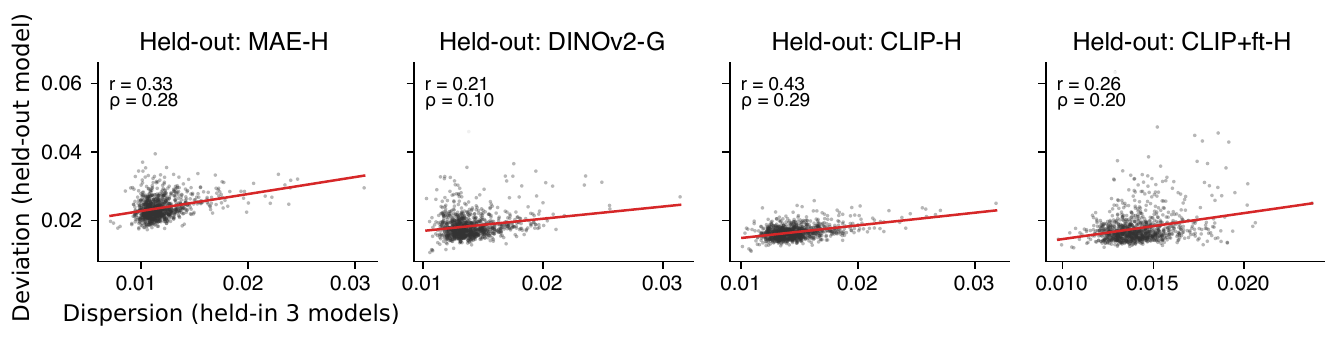}

  \caption{\textbf{Dispersion computed from a subset of the vision models predicts the held-out model's deviation from the consensus}. For each of the four vision models in turn, GPA is re-run on the other three, producing a partial-dispersion score per stimulus (\textit{x}-axis). The held-out model is then projected into the remaining consensus and its per-stimulus deviation from that consensus is measured (\textit{y}-axis). Each point is one WIT stimulus (n = 1024); the red line is the linear fit. Pearson $r$ is significantly positive for all four held-out models: MAE-H $r = 0.33$, $p < 0.001$; DINOv2-G $r = 0.21$, $p <0.001$; CLIP-H $r = 0.43$, $p < 0.001$; CLIP+ft-H $r = 0.26$, $p < 0.001$ ($\rho$ values show Spearman correlations). This suggests that stimulus-level dispersion captures a property shared across model families.}

  \label{fig:S6_heldout_model}

\end{figure}

\end{document}